# Optical valley Hall effect for highly valley-coherent exciton-polaritons in an atomically thin semiconductor

Nils Lundt[1], Lukasz Dusanowski[1], Evgeny Sedov[2,3], Petr Stepanov[4], Mikhail M. Glazov[5], Sebastian Klembt[1], Martin Klaas[1], Johannes Beierlein[1], Ying Qin[6], Sefaattin Tongay[6], Maxime Richard[4], Alexey V. Kavokin[7,8,2], Sven Höfling[1,9] and Christian Schneider[1]

[1]*Technische Physik and Wilhelm-Conrad-Röntgen-Research Center for Complex Material Systems, Universität Würzburg, D-97074 Würzburg, Am Hubland, Germany.*

[2]*Physics and Astronomy School, University of Southampton, Highfield, Southampton, SO171BJ, UK*

[3]*Vladimir State University named after A.G. and N.G. Stoletovs, Gorky str. 87, 600000,Vladimir, Russia*

[4]*Univ. Grenoble Alpes, CNRS, Grenoble INP, Institute Neel, 38000 Grenoble, France*

[5]*Ioffe Institute, Polytechnicheskaya 26, 194021 St. Petersburg, Russia*

[6]*Arizona State University, Tempe, Arizona 85287 USA*

[7]*Westlake University, 18 Shilongshan Road, Hangzhou 310024, Zhejiang Province, China*

[8]*Institute of Natural Sciences, Westlake Institute for Advanced Study, 18 Shilongshan Road, Hangzhou 310024, Zhejiang Province, China*

[9]*SUPA, School of Physics and Astronomy, University of St. Andrews, St. Andrews KY 16 9SS, United Kingdom*

**Spin-orbit coupling is a fundamental mechanism that connects the spin of a charge carrier with its momentum[1]. Likewise, in the optical domain, a synthetic spin-orbit coupling is accessible, for instance, by engineering optical anisotropies in photonic materials[2]. Both, akin, yield the possibility to create devices directly harnessing spin- and polarization as information carriers[3]. Atomically thin layers of transition metal dichalcogenides provide a new material platform which promises intrinsic spin-valley Hall features both for free carriers, two-particle excitations (excitons), as well as for photons[4]. In such materials, the spin of an exciton is closely linked to the high-symmetry point in reciprocal space it emerges from (K and K' valleys)[5,6]. Here, we demonstrate, that spin, and hence valley selective propagation is accessible in an atomically thin layer of MoSe$_2$, which is strongly coupled to a microcavity photon mode. We engineer a wire-like device, where we can clearly trace the flow, and the helicity of exciton-polaritons expanding along a channel. By exciting a coherent superposition of K and K' tagged polaritons, we observe valley selective expansion of the polariton cloud without neither any applied external magnetic fields nor coherent Rayleigh scattering. Unlike**

**the valley Hall effect for TMDC excitons[7], the observed optical valley Hall effect (OVHE)[8] strikingly occurs on a macroscopic scale, and clearly reveals the potential for applications in spin-valley locked photonic devices.**

Spin-valley locking is a striking feature of free charge carriers and excitons emerging in monolayers of transition metal dichalcogenides (TMDCs) [6,9]. It originates form the strong spin-orbit interaction, which arises from the heavy transition metals in TMDCs and the broken inversion symmetry of the crystal lattice. This leads to inverted spin orientations at opposite K points at the corners of the hexagonal Brillouin zone, for both conduction band electrons and valence band holes. As a result, the K and K' valleys can be selectively addressed by σ+ and σ- circular polarized light[10,11], which is referred to as valley-polarization. Likewise, coherent superpositions of both valleys can be excited by linear polarized light, which is referred to as valley coherence. The outstanding control of the valley pseudospin has attracted great interest in exploiting this degree of freedom to encode and process information by manipulating free charge carriers[12] and excitons[7,13,14], which has led to the emerging field of valleytronics[4]. However, exciton spin-valley applications are strongly limited by the depolarization mechanisms due to the strong Coulomb exchange interaction of electrons and holes, as well as by the limited exciton diffusion and propagation lengths.

Thus far, in most experiments, TMDC monolayers were non-resonantly excited several hundreds of millielectronvolt above the ground state, so that excitons were created with a finite center of mass wave vector that deviates from the quasi-momentum of the K/K' valley[15]. In this configuration, the long-range Coulomb exchange interaction of the exciton's constituent electron and hole creates an efficient depolarization mechanism[16,17]. The eigenstates of excitons with their dipole orientation parallel to the wave vector (longitudinal exciton) and with their dipole orientation perpendicular to the wave vector (transverse exciton) are split in energy. This so called L-T splitting increases linearly with the center of mass wave vector, and mixes the K and K' excitons, which ultimately leads to a depolarization of excitons during their energy relaxation[18].

One way to reduce this depolarization effect, is to excite the system quasi-resonantly close to the exciton resonance energy. As a result, excitons are created with small in-plane wave vectors, corresponding to a relatively small L-T splitting and they need to undergo less inelastic scattering events before their radiative decay as compared to non-resonantly excited excitons. This approach has been successively used to measure reasonably high DOCP values in $MoS_2$ (40 %)[19], $WS_2$ (90 %)[20] and $WSe_2$ (40 %)[21] monolayers at cryogenic temperature and DOCP values up to 30 % in a $WS_2$ monolayer a room temperature[22]. However, in $MoSe_2$ only a few percent DOCP have been observed this way[23,24]. To achieve close-to-unity valley polarization, excitons in $WSe_2$ have been resonantly excited in a two photon absorption process[25].

An alternative approach to enhance the valley polarization is the tailoring of the electrodynamic environment of the monolayer. This can be accomplished by integration of a TMDC monolayer into an optical microcavity, resulting in the formation of exciton-polaritons (polaritons)[24,26]. This approach has been used in particular to increase the DOCP in $MoS_2$ and $WS_2$ monolayers at room temperature[27–29] and to induce a significant valley polarization in $MoSe_2$ monolayer[24,30] at cryogenic temperatures. Here, the strong coupling between the optical cavity mode and the monolayer exciton speeds up the energy relaxation and decay dynamics, which is beneficial for preserving the valley polarization[24,27,29]. Moreover, inter-valley scattering is greatly suppressed in the strong coupling regime[24], owing to the strongly reduced disorder scatting of the photon component of the polariton[8]. In addition, polaritons in TMDC monolayers have a much longer diffusion length as compared to excitons in a bare monolayer[31–33], which should be a great advantage with respect to spatial separation of valley tagged excitations.

In this work, we take advantage of these two approaches to enhance valley polarization and valley coherence in a TMDC monolayer. First, we strongly couple excitons in an elongated monolayer $MoSe_2$ to the optical mode of a high quality factor, mechanically assembled microcavity. Next, we excite the exciton-polariton states quasi-resonantly close to their ground state by a two-photon absorption process. Combining both approaches enables us to generate valley polarization and valley coherence of the quasi-particles with circular and linear polarization degrees of more than 90%. The ultimate control over valley polarization and valley coherence is a key to observe the optical valley Hall effect. In this effect the polaritons with opposite valley states propagate in different directions resulting in the valley separation polariton of exciton-polaritons alike spin-up and spin-down electrons in a semiconductor driven by an electric field separate[34]. We evidence this effect by studying the polarization of polaritons expanding along the monolayer. The spin-Hall phenomenology in optical distributed Bragg reflector (DBR) cavities, which is translated to the valley Hall physics in chiral valley monolayers, arises from the splitting of the transvers electric (TE) and transvers magnetic (TM) modes at finite in-plane wave vectors[35]. This is qualitatively similar to the excitons' L-T splitting, but the TE-TM splitting is approximately two orders of magnitude larger[8]. Thus, if excitons strongly couple to a microcavity photon mode and propagate with finite velocities, the emergent polaritons experience a spin-dependent propagation[35]. In our sample, this is manifested in a clearly observed and well controllable spatial separation of valley tagged polaritons, which is a clear-cut feature of the OVHE[8].

**Sample structure and polariton formation**

The studied sample structure is schematically depicted in Fig. 1a. The mechanically assembled microcavity is built by transferring a $MoSe_2$ monolayer with a dry-gel method[36] onto a $SiO_2/TiO_2$ bottom DBR (10 pairs, stop band center at 750 nm). The monolayer was mechanically exfoliated from a bulk crystal. Subsequently, we spin-coated a 126 nm thick poly-methyl-methacrylate (PMMA) buffer layer. Finally, a piece of a separate $SiO_2/TiO_2$ DBR with lateral dimensions of a few tens of μm (8.5 pairs, stop band center at 750 nm) is mechanically peeled off its substrate. This piece is transferred onto the buffer layer with the same dry-gel method[36]. Van der Waals forces are sufficiently strong to keep the top DBR in place. We found that such microcavities demonstrate Q-factors up to 4600 (see appendix). Fig. 1b shows a microscope image of the final structure (top view), where the MoSe$_2$ monolayer (8 x 40 µm) is marked in blue. A photoluminescence (PL) spectrum of the monolayer capped with PMMA was taken at 5 K (Fig. 1c) before the microcavity was completed with the top DBR. The monolayer exhibits a neutral exciton resonance at 745.5 nm/1.663 eV and a charged exciton (trion) resonance at 759.8 nm/1.632 eV. Figure 1c also shows a reflectivity spectrum of the uncoupled cavity mode at 747.3 nm/ 1.659 eV, measured in the close to the monolayer position.

First, we probe the in-plane-momentum-resolved PL of the full structure at the monolayer position at 5 K under continuous-wave excitation with a titanium-sapphire laser (Msquared Solstis) at 740 nm. Neither the cavity mode, nor exciton or trion mode can be observed but a new mode appears with a ground state at 761.4 nm/1.628 eV, as presented in Fig. 1d. This mode has a parabolic dispersion at small in-plane momentum components parallel to the long axis of the monolayer. In contrast, if the dispersion is measured perpendicular to the monolayer extension, discrete levels appear in the PL spectrum as depicted in Fig. 1e. The mode splitting of 2.7 meV between the fundamental and first excited mode is very consistent with an optical mode confinement induced by the monolayer width of 8 µm. We attribute this new mode to the lower polariton branch of the strongly coupled exciton-cavity system. The upper polariton branch, and likewise, lower polariton states with large in-plane wave vectors, are typically not observed in photoluminescence spectra acquired under non-resonant pumping, since the thermal population of polaritons can be very low at cryogenic temperatures[37]. In order to provide further evidence for the formation of exciton-polaritons, i.e. the strong light-matter coupling regime, we have carried out white-light reflection measurements at various temperatures (see appendix). Based on these measurements, we have modelled the presented lower polariton dispersion by a two-coupled-oscillator model, which yield a normal mode splitting of 46 meV and an exciton fraction of 26 % in the ground state (see appendix). The two-coupled-oscillator model is also presented as an overlay to the measured dispersion in Fig. 1d.

**Polariton valley polarization and valley coherence**

TMDC monolayers can be excited under two-photon absorption[25], which either originates from a second harmonic generation (SHG) process or a two-photon interband transition[38,39]. This strong non-linear response is transferred to our strongly coupled system[40] and thus provides an ideal setting to study polaritons under nearly resonant excitation conditions. This allows us to directly address polaritons arising from specific valleys of the embedded crystals via quasi-resonant chiral two-photon absorption. The clean, high-fidelity excitation of valley polaritons is compulsory for studies of exciton-polaritons in valleytronic experiments. Here, we excite our system at 5 K with 2 ps laser pulses at 1518 nm (Coherence Mira-OPO system, 82 MHz repetition rate) with σ$^+$ and σ$^-$ circular polarization and analyze the polarization of the emitted light (Fig 2a). The measured spectra for σ$^+$ and σ$^-$ excitations are plotted in Figs. 2b and c, respectively. We observe that the emitted signal is strongly polarized with an opposite helicity as compared to the polarization of excitation. Observing a counter-rotating signal is fully consistent with SHG selection rules for crystals with broken inversion symmetry and a three-fold rotation symmetry[41], and likewise for two-photon active interband transitions[25,38]. In Figs. 2b and 2c, we plot the degree of circular polarization of the emission across the acquired spectrum, including genuine second harmonic signals and luminescence following the two-photon absorption process (more details on the distinction of SHG vs. two-photon-induced photoluminescence (2P-PL) are discussed in the apendix). We find that a valley polarization of more than 90%, as demonstrated by the corresponding DOCP, is preserved even at energies that are notably below the second harmonic resonance, which is characteristic for 2P-PL[25,39]. The outstandingly high valley polarization is explained by two factors: First, the excited polariton states are situated close to the ground state, and therefore their in-plane wave vectors are low (< 1.5 μm$^{-1}$). Hence, the TE-TM splitting of the cavity and the renormalizations of the exciton oscillator strengths in L and T polarizations, which are the two main mechanism of the depolarization, are small as compared to the case of non-resonant excitation. Second, the polariton lifetime is shortened to about 260 fs by coupling to the cavity (see appendix) as compared to pure excitons in $MoSe_2$ monolayer (390 fs[42]). As assessed in Refs.[8,24], the strong coupling conditions mitigate the influence of strong exciton localization which leads to depolarization (via broadening in k-space), and enable valley relaxation timescales that are extended by a factor of approximately 20. In particular, the negative detuning of exciton and cavity photon mode is crucial for such suppression of disorder scattering[8]. Thus, a strong increase in the degree of circular polarization can be expected in our resonantly driven cavity polariton system. The resulting value of DOCP is governed by the interplay between radiative decay time $\tau$ and spin-valley relaxation time $\tau_s$ and can be approximated by $\rho_{circ} = \rho_0/(1 + \frac{\tau}{\tau_s})$[15,28], where $\rho_0$ is the initial polarization (close to 100%). The spin-valley relaxation time for the pure $MoSe_2$ monolayer exciton is 150 fs[24,42], resulting in a polariton spin-valley relaxation time of 3 ps[24]. With this, we calculate a valley polarization of 92%, which is in excellent agreement with our experimental observations.

While valley-polaritons can be generated by exciting with circular polarized light as shown above, excitation with linearly polarized light may induce a coherent superposition of valley polaritons. Valley coherence up to 80% has been observed in bare monolayers of $MoS_2$[43], $WS_2$[22] and $WSe_2$[44], but has not been measured in $MoSe_2$ monolayers. For our valley polaritons, the degree of linear polarization (DOLP) $\rho_{lin}$ of the decaying polariton field provides a direct measure of the dephasing processes of the polariton pseudospin vector on the equator on the Bloch sphere (Fig. 2d). We probe our system by exciting at 1514 nm (757 nm) with linearly polarized light in the X- and Y-basis, where the X axis is aligned with one of the crystal axis and the Y axis is perpendicular to the X axis. The emitted signal is subsequently measured in the same basis and the DOLP was determined by $\rho_{lin} = \frac{I(X)-I(Y)}{I(X)+I(Y)}$. Figures 2e and 2f show the acquired spectra, along with the calculated DOLP values as a function of the detection wavelength. The polarization of the resonantly scattered fraction of the signal (the SHG part) is expected to be fully governed by the crystal symmetry and yields no information about the valley coherence of the polaritons. However, it is remarkable that a strong linear polarization is maintained in the broad 2P-PL emission of the polariton resonances, even 10 meV below the energy of the SHG signal, with a DOLP clearly exceeding 90% close to the polariton ground state. Thus, the DOLP clearly reflects the valley coherence.

The phase of the coherent valley superposition can be addressed by changing the direction of the linear polarization of the pump beam. In Fig. 2f, we show that 2P-PL retains the direction of the linear polarization of the pump oriented along the Y-direction, and likewise allows us to extract large degrees of valley coherence exceeding 90 % even 10 meV below the SHG frequency. In contrast, the DOLP signal is close to zero or even negative around the SHG frequency, due to the specific SHG selection rules in TMDC monolayer[38].

While near-unity valley polarization has been observed on bare $WSe_2$ monolayers[25], the demonstrated, unprecedented degrees of valley coherence, in particular in a $MoSe_2$ monolayer, clearly outline the potential of strong light-matter coupling for generation and manipulation of coherent valley superposition states of hybrid light-matter quasiparticles.

**Optical Valley Hall Effect**

In contrast to TMDC excitons, which have a diffusion length of a few hundred nanometers in high quality samples[32,33], TMDC based polaritons are expected to expand over significantly larger distances (on the order of 10 µm) even in the linear, non-ballistic regime due to their small effective masses[31]. Figure 3a depicts the spatial intensity distribution of near-resonantly injected polaritons on the monolayer at 5 K. From this intensity distribution, we can infer that our polaritons propagate by approximately 3.6 µm before they decay (see appendix). It should be noted that an excitation slightly above the LPB (in this case 10 meV) is essential to create polaritons with finite wave vectors and velocities. Such long propagation lengths allow us to investigate the valley dependent polariton propagation upon valley coherent initialization of the system. This is recorded via spatial- and polarization-resolved luminescence. Here, the real space intensity distribution was measured in $\sigma^+$ and $\sigma^-$ basis at 5 K and the spatial-resolved DOCP was deduced. Figure 3b depicts a DOCP distribution recorded under linearly polarized two-photon excitation at 1514 nm. The initial polarization angle was -75° with respect to the long monolayer axis. Clearly, the DOCP distribution shows two domains with a left-right separation that is slightly rotated clockwise and features an oscillating pattern along the vertical stripe direction. As we rotate the initial polarization orientation by -45° (Fig. 3c), we observe a dramatic change in the polarization pattern with a vertical domain separation and weak $\sigma^+$ regions along the right monolayer edge. We note that similar patterns have been predicted[35] and observed[45,46] as spin domains of exciton-polaritons in GaAs QWs embedded in microcavities, known as the optical spin hall effect (OSHE)[35,45]. In contrast to the GaAs case, the spin and valley indices cannot be treated separately in TMDC monolayer excitons, and consequently, the valley-dependent expansion of polaritons leads to the emergence of the OVHE[8].

Based on the physics of the OVHE effect, the intensity and DOCP distributions were modelled for the measured monolayer geometry and the experimental conditions. We describe our system by solving the generalized Gross-Pitaevskii equation for the two-component wavefunction $\Psi_S(\mathbf{r},t)$

$$i\hbar \partial_t \Psi_\sigma = \left[\hat{E} + V(\mathbf{r})\right]\Psi_\sigma + \frac{\Delta}{2}\left(\hat{k}_x - \sigma i \hat{k}_y\right)^2 \Psi_{-\sigma} + i\frac{\hbar}{2}\left(R n_R^\sigma - \gamma\right)\Psi_\sigma + P_\sigma(\mathbf{r})$$

coupled to the rate equation for the density of the spin-resolved reservoir of incoherent excitons:

$$\partial_t n_R^S = Z|P_S(\mathbf{r})|^2 - \left[g_R + R|\Psi_S|^2\right]n_R^S,$$

where $S$ describes either spin-up (+) or spin-down (-) states. $\hat{E}$ is the polariton kinetic energy operator, which reproduces the non-parabolic polariton dispersion in Fig. 1d observed experimentally and fitted using the two-coupled-oscillator model (appendix). $V(\mathbf{r},t) = V_0(\mathbf{r}) + a|\Psi_S|^2 + a_R n_R^S$ is the effective potential experienced by polaritons. It consists of the stationary confinement potential across the sample $V_0(\mathbf{r})$ and the blueshift induced by polariton-polariton interactions within the condensate and interactions of polaritons with excitons in the incoherent reservoir. $a$ and $a_R$ are the corresponding interaction constants. $\Delta$ is the TE-TM splitting constant, $\hat{\mathbf{k}} = (\hat{k}_x, \hat{k}_y) = (-i\partial_x, -i\partial_y)$ is the quasimomentum operator. The polariton condensate is excited by the non-homogeneous resonant optical pump $P_S(\mathbf{r})$. To take into account the inevitable appearance of the reservoir of incoherent excitons under the resonant pumping, we introduce the term $Z|P_S(\mathbf{r})|^2$; $Z$ is the dimensional reservoir response constant[47]. $R$ describes the stimulated scattering rate from the reservoir to the ground state. $g$ and $g_R$ are the decay rates of polaritons and reservoir excitons, respectively.

For the simulation, we take values of the parameters estimated from the experiment. We take the decay rates as $g = 1/260$ fs$^{-1}$, $g_R = 1/390$ fs$^{-1}$. The pump energy is resonant to the polariton energy at $k = 1.5\,\mu m^{-1}$. The LT-splitting is taken as $\Delta k^2 = 0.75$ meV typical for such kind of structures[30]. The stimulated scattering rate is $R = 0.1$ meV $\mu m^2$, the interaction constants are $a = a_R / 3 = 0.5$ meV $\mu m^2$.

The simulated intensity distribution, plotted in Fig. 3d, shows a polariton propagation tail along the monolayer similar to the experimentally observed one. The simulated DOCP distributions, calculated as $(|\Psi_+|^2 - |\Psi_-|^2)/(|\Psi_+|^2 + |\Psi_-|^2)$, for the same initial linear polarization orientations that were used in the experiments are plotted in Fig. 3e and f, respectively.

In the case of the -75° orientation, the simulation nicely reproduces the valley separation. As in the experiment, the DOCP is increasing towards the flake edges. Polaritons that decay further away from the excitation spot are associated with higher in-plane wave vectors, which in turn provide higher splittings and stronger effective magnetic fields. Consequently, the precession of the pseudo-spin, which is equivalent to the polarization Stokes vector, is more pronounced leading to a larger valley polarization. Moreover, the wavy domain separation line is also clearly seen in the simulation. This is explained with interfering polaritons that are reflected at the monolayer edges. The simulation of the rotated initial polarization in Fig. 3f is again in good qualitative agreement with our experimental data. The main, bottom section of the monolayer is strongly dominated by a $\sigma^-$ domain with small $\sigma^+$ regions at the right monolayer edge.

The very good agreement of our experimental data with the theoretical modelling allows us to interpret our experimental data as the first manifestation of the OVHE in a TMDC exciton-polariton system. We note, that our experimentally observed valley polarized domains yield DOCP up to 80%, which itself is a remarkable consequence of the strong valley polarization and valley coherence, which is preserved by our experimental approach.

**Conclusion**

We have demonstrated that near-unity valley polarization and valley coherence can be achieved under near-resonant two-photon excitation by integrating a TMDC monolayer into a high-Q microcavity. Such high degrees can be achieved since strong coupling conditions strongly decrease the radiative lifetime and the depolarization time is significantly prolonged at states such close to the polariton ground state. It is remarkable that this mechanism yield such high degree even in $MoSe_2$ monolayers, which is known exhibit a rather poor DOCP as bare monolayer[24]. It has been argued that the presence of a low-lying dark exciton state in the tungsten based TMDC materials, which is not subject to the MSS mechanism, is beneficial for the observation of large degrees of valley polarization and coherence[48]. Thus, we suspect that similar experiments, with even more pronounced degrees of polarization can be conceived based on high quality, tungsten based TMDC layers, potentially also a elevated temperatures.   Under these conditions, we have observed valley dependent, optically controllable propagation of exciton-polaritons in a high quality factor microcavity, containing a channel-like TMDC monolayer structure. By taking advantage of non-linear, quasi-resonant spectroscopy in the strong coupling regime, we observed highest degrees of valley polarization and valley coherence in our system, which reflects the potential of coherent light-matter coupling. Clear signatures of polarization domains become visible in the expansion of polaritons, observed in the valley-coherent excitation scheme. This is a clear-cut signature of the optical valley Hall effect, which allows to observe the interplay of spin, valley, and momentum of our quasi-particles, driven by internal pseudo-magnetic fields. While the OVHE propagating polaritons may be utilized in valleytronic on-chip applications, the interplay of valley coherent superpositions and photonic spin-orbit coupling could be exploited to harness valley-path entanglement phenomena of propagating quantum wavepackets of polaritons. Likewise, by carefully preparing monolayers with imprinted super-potentials for excitons, similar experiments could be conceived to demonstrate topologically non-trivial expanding polaritons in a similar framework[49], and even pave the way towards topological lasers in the strong coupling regime5[50] based on TMDC crystals.

## Acknowledgements

C.S. acknowledges support by the ERC (Project unLiMIt-2D). The Würzburg group acknowledges support by the State of Bavaria. A.V.K. acknowledges the support from Westlake University (Project No. 041020100118). E.S.S. acknowledges support from the RFBR Grant No. 17-52-10006. S.K. acknowledges support by the EU (Marie Curie Project TOPOPOLIS). Q.Y. and S.T. acknowledge funding from NSF DMR-1838443 and DMR-1552220. M.M.G. acknowledges partial support from RFBR Project 17-02-00383.

Correspondence and requests for materials should be addressed to Christian Schneider (christian.schneider@physik.uni-wuerzburg.de).

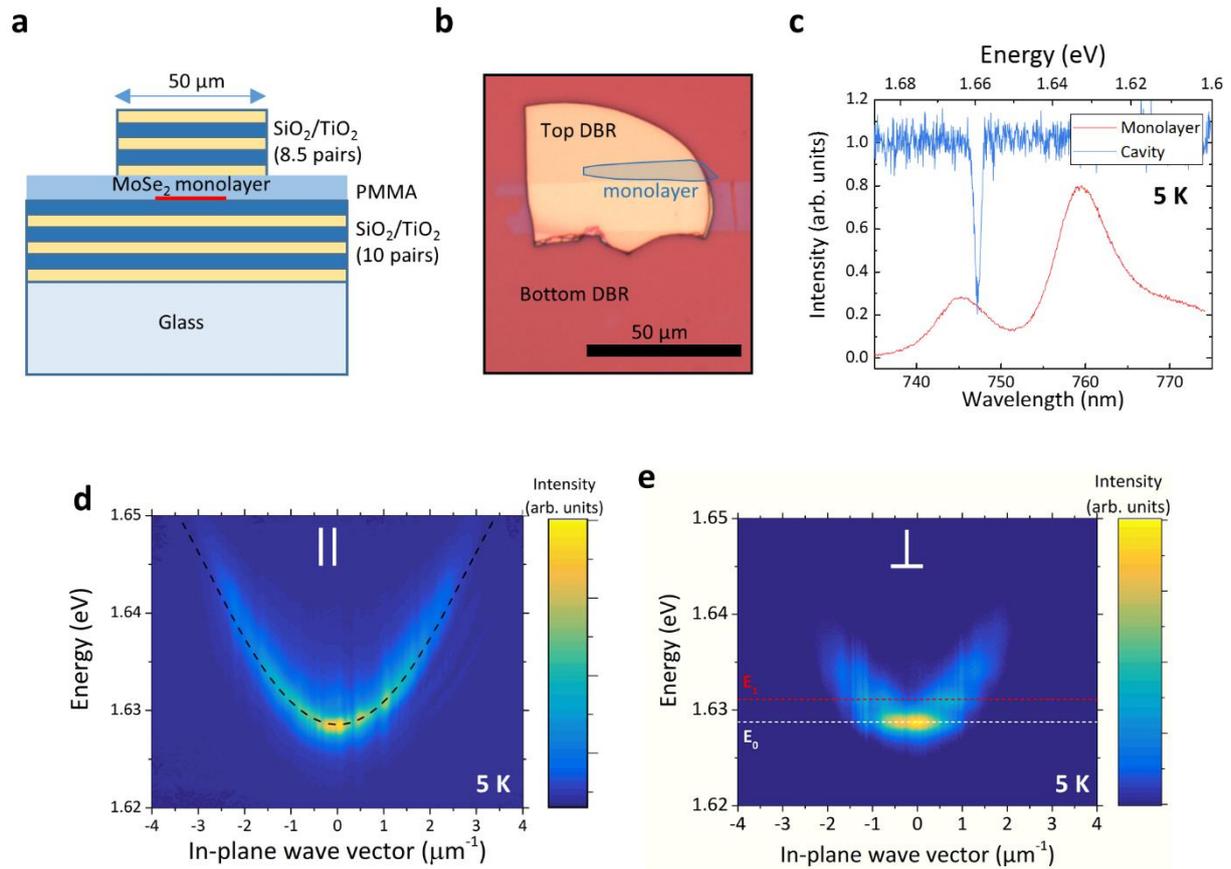

Fig. 1 | Sample design and characterization. a, Sketch of the mechanically assembled microcavity. b, Optical Image of the cavity, indicating the elongated monolayer, which provides the channel geometry. c, Photoluminescence measurement of the untreated monolayer and the optical empty-cavity resonance at 5 K. d, Momentum-resolved photoluminescence spectrum recorded at 5 K along the stripe-direction, as well as perpendicular to the stripe (e). In d, the measurement is overlaid with a two-coupled oscillator model. In e, the mode discretization, which arises from the finite monolayer width of 8 µm is marked for the first (black) and second (red) discretized state.

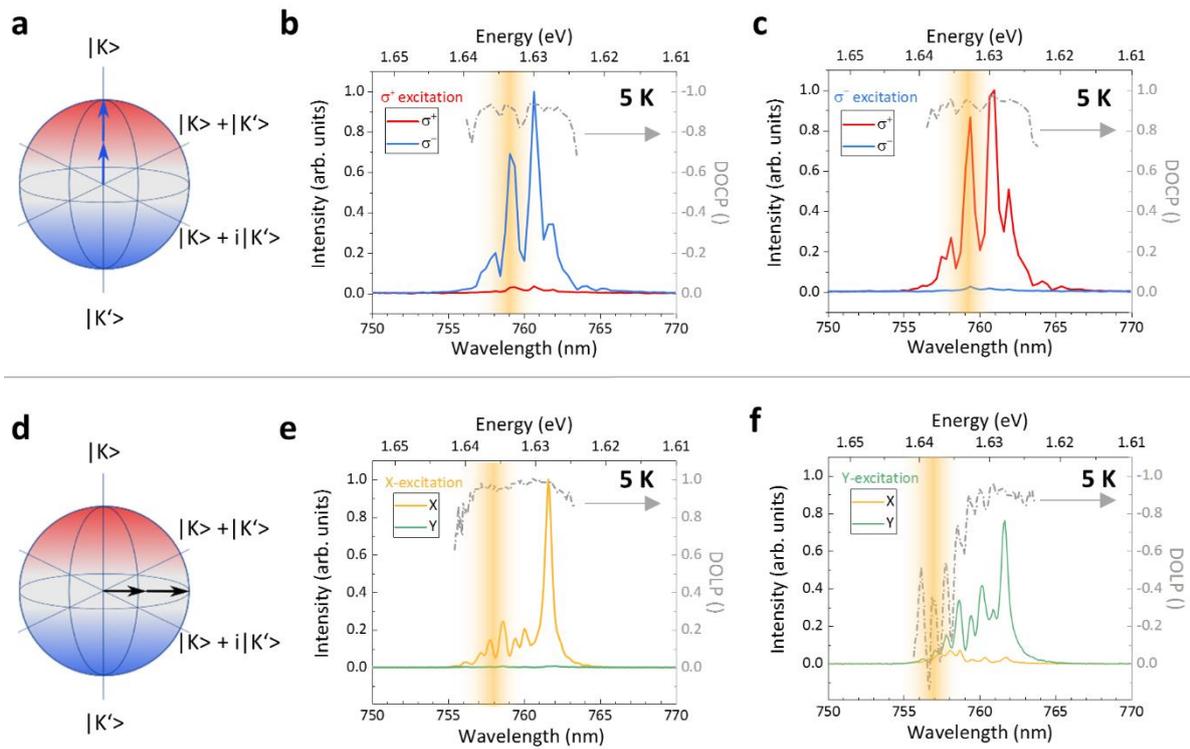

*Fig. 2 | **Valley polarization and coherence of polaritons.** a) Schematic drawing of the polariton pseudo-spin, which is generated at K/ K' by quasi-resonant two-photon absorption. b) Polarization-resolved emission spectrum of polaritons, excited at 5 K by a $\sigma^+$ polarized near-infrared laser ($\sigma^-$ polarized laser depicted in c)). The degrees of circular polarization of the emitted signals reveal a strongly counter-polarized emission. The energy of the second harmonic resonance is indicated by the orange area. d) Schematic drawing of the excitation of a valley coherent state, generated by quasi-resonant two-photon absorption. e) Polarization-resolved emission spectrum of polaritons, excited at 5 K with a linearly polarized near-infrared laser in X-basis (Y-basis, linear polarized injection is depicted in f)). The emitted light is strongly co-polarized. The energy of the second harmonic generation resonance is indicated by the orange area.*

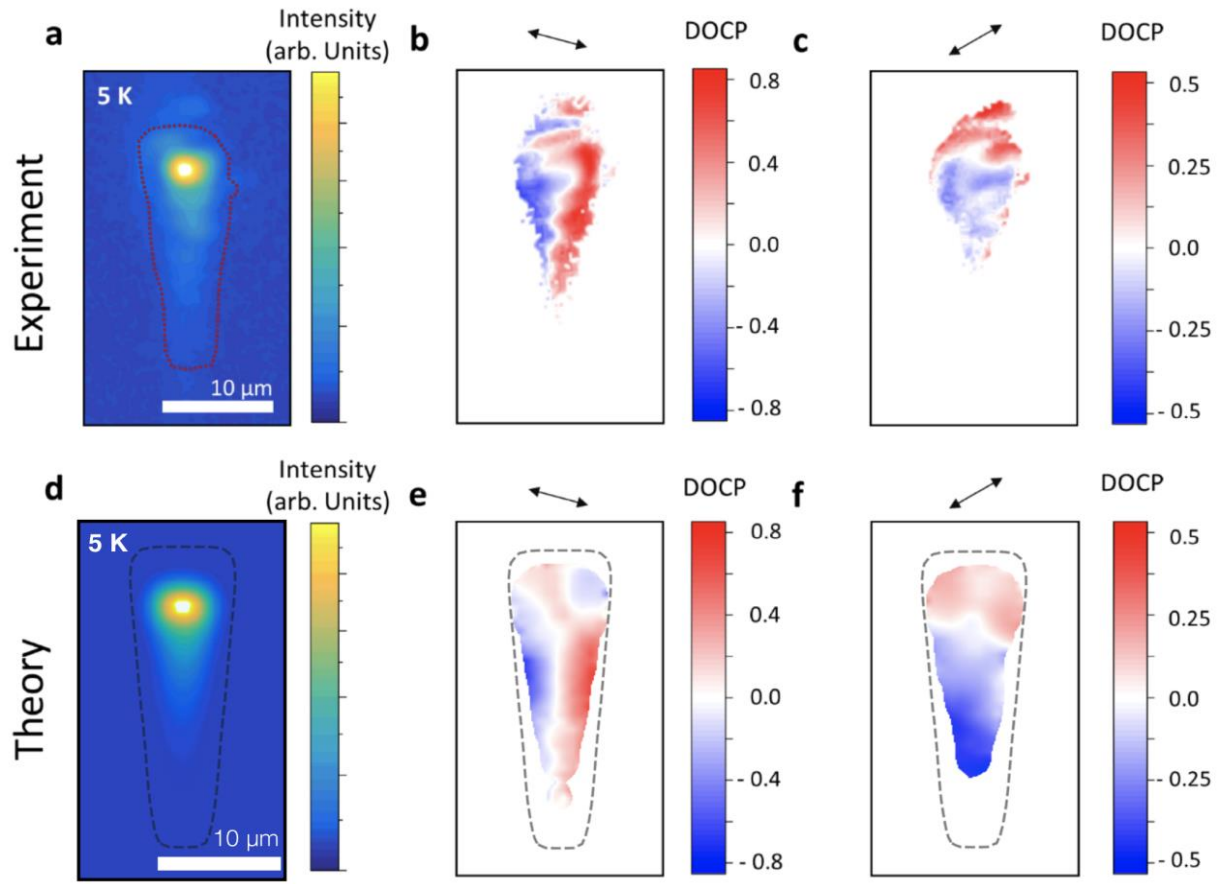

*Fig. 3 | **Optical Valley Hall Effect.*** *a) PL intensity distribution along the monolayer, observed under two-photon excitation at 1514 nm, where the white point marks the excitation spot on the monolayer. The monolayer edges are indicated by the dotted line. b) Spatially resolved DOCP distribution across the monolayer. The polarization orientation of excitation light (-75° with respect to the long monolayer axis) is indicated by the double-headed arrow. (c) DOCP distribution across the monolayer with a polarization orientation that was rotated by -45° (-120° with respect to the long monolayer axis). d-f) Simulated intensity and DOCP distributions, each corresponding to the figure above.*

# Appendix 1

In Figure 1d of the main text, we plot the dispersion relation of the lower polariton branch, acquired under non-resonant pumping, together with the coupled oscillator fit to the data. We will now provide evidence, that the luminescence occurs in the strong coupling regime. In Figure A1a) we reproduce the spectrum plotted in fig 1d in the main text. In order to fit the observed lower polariton branch, we applied the conventional two-coupled oscillator model. The cavity energy for this model was taken from the anti-crossing experiment presented in fig A2 (1.642 eV), hence it is not a free fitting parameter. For the best fitting result, the exciton energy, which entered the model, was assumed to be 5 meV above the value measured in Figure 1c in the main text. The slight difference can be explained by an uneven influence of the PMMA layer and/or slight strain conditions. Still, it is within the inhomogeneous broadening. The calculated dispersion relation for the lower branch, shown in figure A1a and figure 1d (main text) yields the best match to the experimental data for a normal mode splitting of 46 meV. The Hopfield coefficients of the LPB are plotted in fig. A1c as a function of the in-plane wave vector, showing that the lower polariton is predominantly photonic with a cavity fraction of 74 %.

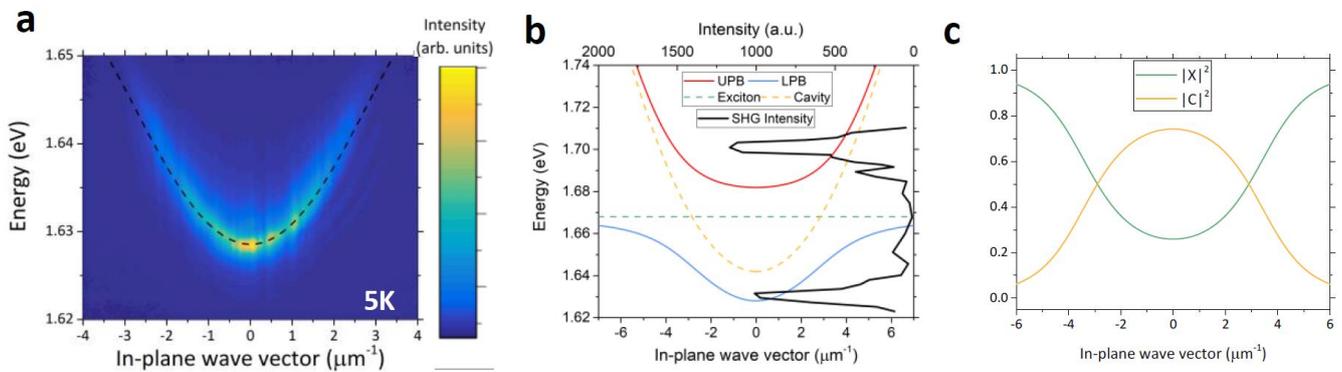

*Figure A1: a) Photoluminescence from the lower polariton state (non-resonant pumping). The black, dashed line is the result of the coupled oscillator fit. b) Plot of the full two-coupled-oscillator model for the measured polariton: dispersion relation for upper and lower polariton modes are shown in solid blue and red lines. The uncoupled exciton (red, dashed line) and cavity (black, dashed line) are also indicated here. Along with the coupled oscillator fit, we depict the result of a scanning second harmonic generation experiment. We find, that the SHG signal corresponds with the predicted polariton states of the coupled harmonic oscillator mode, and furthermore the SHG intensity scales with the photonic Hopfield coefficient of the LPB and UPB states b) Exciton- and Photon Hopfield coefficients for the lower polariton branch.*

In figure A1b) we plot the complete coupled oscillator dispersion relation (corresponding with the fitting result shown in figure A1a). Along with the modelled dispersion relation, we plot the result of a scanning second harmonic generation (SHG) experiment. Here, we tuned our linearly polarized near infrared laser (2 ps, 82 MHz) between 1450 and 1528 nm, and recorded the second harmonic signal generated from the polariton resonances. The SHG intensity generated from the sample is plotted alongside with the calculated dispersion relation in Fig S2b as a function of twice the pump energy.

First, we would like to state, while the broken inversion symmetry of the TMDC crystal enforces SHG, similar effects have previously been observed in GaAs microcavities in the strong coupling regime (ref. 40, main text).

In our experiment, we find a strong enhancement of the SHG signal when the laser is scanned over the lower polariton branch. In particular, the SHG signal acquires a significant intensity for polariton states with high photonic Hopfield coefficients, near the ground state. This correlation of SHG and the photonic Hopfield coefficient has been observed in previous studies (see Ref 1.), where the SHG intensity was found to be most intense for polaritonic resonances with photonic fractions > 50 % . For frequencies around the anti-crossing region, the SHG signal is not observable in our experiment, but importantly, we observe a second peak in the SHG signal where our coupled oscillator model predicts the upper polariton branch. While weak SHG signals indeed can be observed starting from the lowest energy states of the UPB (with large excitonic fraction), again, the SHG intensity increases for higher energy states in the UPB with larger photonic fraction. This, again, reflects the direct correlation of SHG intensity and the photon Hopfield coefficient previously found in GaAs microcavities in the strong coupling regime.

To summarize, our measured dispersion relation of the lower polariton branch can be quantitatively reproduced by a coupled oscillator dispersion, and at the same time, we observe significant SHG intensity when we scan our infrared laser through the resonances of the LPB as well as the predicted UPB, showing the expected dependency on the photonic Hopfield coefficient. Thus, we believe that our data provide direct evidence for the strong coupling regime in our device.

**Appendix 2: the Strong coupling regime in our microcavity: Part II**

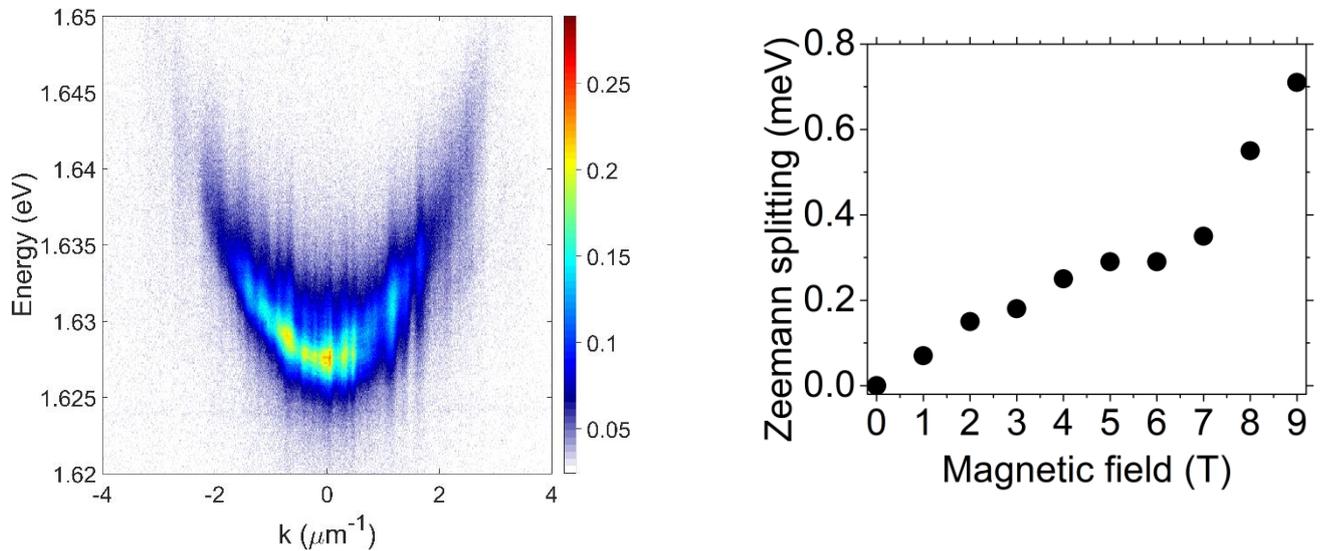

*Figure A2: a) Photoluminescence from the lower polariton state (non-resonant pumping) in the magnetic cryostat. b) Extracted energy splitting of the polariton ground state as a function of applied magnetic field. The extracted Zeeman-Splitting evidences the matter-contribution to our quasi-particles, and thus proves strong coupling conditions in our device.*

Further evidence for the presence of strong coupling conditions in our microcavity is reflected by its response to an external magnetic field: As a result of coherent light-matter coupling, exciton-polaritons sensible react to magnetic fields applied in the Faraday geometry, and exhibit a characteristic Zeeman-splitting. For exciton-polaritons emerging from TMDC crystals, the diamagnetic contribution can usually be safely neglected for magnetic fields < 30 T [2] .Indeed, the Zeeman-splitting for TMDC-based exciton polaritons has been recently observed [3], in closest analogy to experiments conducted on III-V based microcavities [4]. Thus, we have studied our microcavity in the presence of an external magnetic field up to 9T, and extracted the circular polarization splitting of the polariton ground state (figure A2a) as a function of the magnetic field strength. We observe a clear emergence of the polaritonic Zeeman-effect (fig. A2b), which can only be explained by the presence of strong coupling conditions in our device. We note, that a full quantitative treatment of the effect is given in ref 3.

**Appendix 3: Evidence for the Strong coupling regime in our microcavity: Part III**

Since the upper polariton branch is not well-observable under non-resonant optical injection (photoluminescence), in addition to the SHG experiment in S2, a white light reflectivity measurement was carried out, which should reveal the upper polariton branch as its absorption does not require the population of the UPB. Figure A3a shows reflectivity spectra recorded at various temperatures from 10 K to 170 K. The observed resonances show a distinct anti-crossing behavior for a crossing point at about 125 K. The observed normal mode splitting is 30 (+/- 2) meV at this temperature.

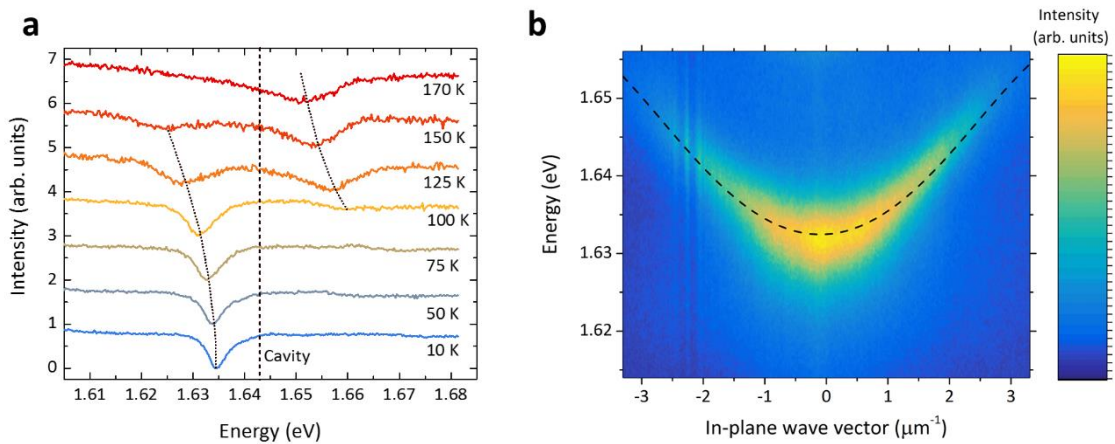

*Figure A3: a) Reflectivity spectra of the full system at various temperatures. The exciton tuned in the temperature series, revealing a distinct anti-crossing behavior. Upper and lower polariton branch as well as the cavity are indicated as guide to the eye. b) Lower polariton dispersion relation measured at 5 K the in photoluminescence. The corresponding two-coupled-oscillator model is indicated by the black-dashed line.*

The extracted cavity resonance is 17 meV below the measured empty cavity, presented in Figure 1c in the main text. It should be noted that the latter measurement was taken 20 – 30 µm away from the monolayer positon. Thus, slight fluctuations of the PMMA thickness and also the real part of the monolayer refractive index explain the shifted cavity resonance. Figure S4b shows the dispersion of the lower polariton branch measured at 5 K in photoluminescence, and the according coupled oscillator fit (using the same parameters as extracted in Fig.A2a). Best agreement between theory and experiment is achieved for a normal mode splitting of 34 meV. This is explained with an increasing exciton oscillator strength towards lower temperature, which in turn increases the normal mode splitting. This observation is in quantitative agreement with the temperature dependent normal mode splitting simulated in reference[5]. It should be noted that the measurements presented in fig S4a and fig A3b, were acquired several months after the PL dispersions presented in Figure 1 in the main text and the study in Fig A1. Here, we have extracted a slightly reduced normal mode splitting (34 meV), as compared to the data analysis corresponding with Fig 1 (46 meV, see also appendix 2). We attribute this difference either to some slight aging effects over such long period of time, or a slight modification of the charging state, which has been recently been shown to directly influences the coupling strength on such a quantitative level [6,7].

To summarize, we could fully map out the full anticrossing curve of the two polariton branches in white-light reflection measurements on our device, adding further direct evidence for the presence of strong coupling conditions in our sample.

**Appendix 4: SHG on MoSe$_2$ cavity polaritons**

In order to study the two photon absorption process in our system in more detail, we excited the structure slightly above the ground state with a laser wavelength of 1514 nm, yielding a SHG signal at 757 nm with a linewidth of 1.5 nm/3.3 meV Figure A4 shows the resulting emission spectrum acquired under such excitation. Besides the SHG emission peak at 757 nm, the spectrum is dominated by the emission from resonances that have lower energies than the SHG signal. In fact, the lowest lying resonance coincides with the ground state of the lower polariton branch. This indicates that exciton-polaritons, which are resonantly excited via two-photon absorption, relax in our system towards their ground-state quite efficiently. We believe that luminescence from these states is a result of a two-photon absorption interband transition since SHG should only be observable at double the excitation energy. In principle, such a transition is spin-forbidden for a 1s exciton (and the corresponding exciton-polariton), however the intermixing with 2p states allows a two-photon interband transition into this state[9,10]. Following reference[11], we therefore refer to these emission features as two-photon-induced photoluminescence (2P-PL). The discretization of the 2P-PL stems for the polariton mode quantization, in the direction perpendicular to the monolayer extension as presented in figure 1f in the main text. The non-linear nature of the two photon absorption process is directly reflected in figure A4b, where we plot the SHG emission intensity (excluding 2P-PL) at double the excitation energy as a function of the excitation power in a double-logarithmic scale. The extracted power law coefficient of 1.997 clearly confirms the two-photon absorption nature of the process, and unambiguously demonstrates the highly non-linear properties of our strongly coupled device. We further study the emission intensity at double the excitation energy (orange marked feature in figure A4a) as a function of the orientation of the linearly polarized excitation light. As previously shown[12], in case of experiments carried out on bare monolayers, we observe an intensity modulation with a 60°-periodicity which stems from the SHG selection rules for the three-fold crystal symmetry.

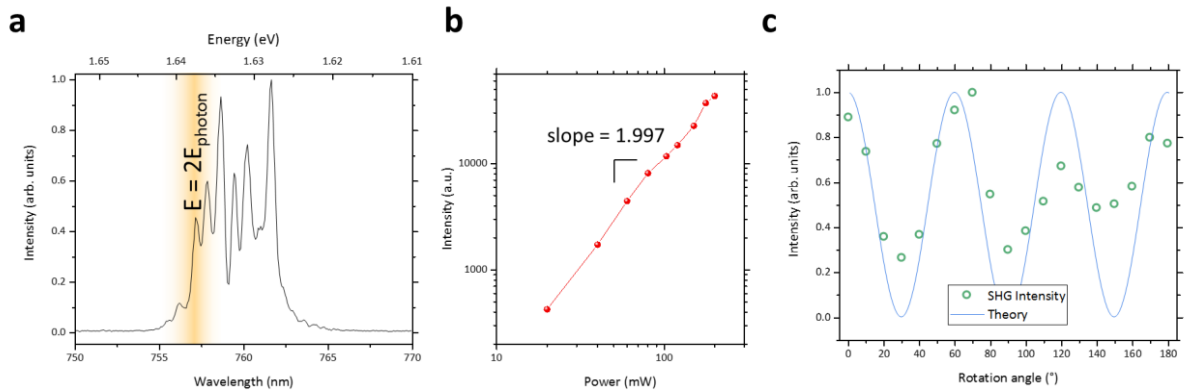

*Figure A4: Two-photon excitation: (a) Emission spectrum under slightly off-resonant excitation at 1514 nm, showing emission features at the SHG energy (marked in orange), but also at lower energies. (b) Double-logarithmic plot of the SHG intensity (intensity from the SHG energy feature only) versus excitation power. (c) Dependence of the SHG intensity on the orientation of the exciting lights linear polarization. (b and c) error bars correspond to symbol size.*

## Appendix 5: Radiative lifetime of the exciton-polaritons

The radiative polariton lifetime $\tau_{polariton}$ can be estimated based on the exciton lifetime $\tau_{exciton}$ and cavity lifetime $\tau_{cavity}$. The lifetime is calculated according to[13]:

$$\frac{1}{\tau_{polariton}} = |X|^2 \frac{1}{\tau_{exciton}} + |C|^2 \frac{1}{\tau_{cavity}} \quad (1)$$

Where $|X|^2$ (0.42) and $|C|^2$ (0.58) are the exciton and cavity fractions of the polariton, respectively (taken from S3). The exciton lifetime $\tau_{exciton}$ is estimated to be 390 fs[14] and the cavity lifetime $\tau_{cavity}$ is calculated according to

$$\tau_{cavity} = \frac{h}{4\pi \Delta E} \quad (2)$$

Where h is Planck's constant and $\Delta E$ is the cavity linewidth. This yields 170 fs for $\tau_{cavity}$ with a $\Delta E$ of 1.54 meV (equivalent to a Q factor of 1030), taken form the reflectivity spectrum in figure 1c in the main text. Finally, a polariton radiative lifetime of 263 fs is extracted.

## Appendix 6: Polariton propagation length

An intensity profile was taken from the intensity distribution presented in figure A6a (see also Fig. 3a in the main text, profile cut is marked by the orange line), which is plotted in figure A6b. This profile was fitted with a convolution of the Gaussian shaped focus profile (FWHM of 2.5 µm) and an exponentially decaying function. This fit yields a decay constant of 3.6 µm, which is taken as an estimate of the polariton propagation length.

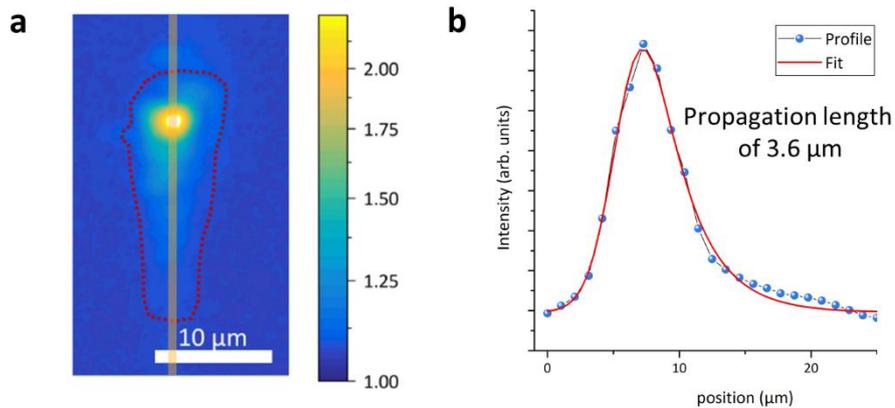

*Figure A6: Polariton diffusion length: a) intensity distribution across the monolayer. The line of the profile that is presented in b) is marked in orange. The profile was fit with a convolution of a Gaussian shaped foucs with an exponential decay function to extract a polariton propagation length of 3.6 µm.*

**Appendix 7: Details on the numerical modelling**

To take the finite size of the TMD monolayer stripe into account and to simulate the confinement of polaritons both along and across the stripe, we take the effective potential $V(\mathbf{r})$ in the form

$$V(\mathbf{r}) \propto 1 - e^{-[2(x-x_0)/w_x]^{12}} e^{-[2y/(1-zx)w_y]^{6}}, \qquad (3)$$

where $w_x$ and $w_y$ are the length and the width of the stripe, $x_0$ is the distance along the x axis with respect to the position of the pump spot. The factor $(1 - zx)$ describes the narrowing of the stripe. The smooth super-Gaussian shape of the potential is chosen to avoid possible numerical issues in the simulations caused by an abrupt variation of the potential landscape, see e.g. Ref.[15].

The pump is taken in the Gaussian form

$$f(\mathbf{r}) \mu \, e^{-[x^2+(y-y_1)^2]/w_p^2} e^{i(k_p x - w_p t)}, \qquad (4)$$

where $w_p$ is the pump width, $\hbar\omega_p$ and $k_p$ are the pump energy and the pump wave number. The parameter is used to take into account the slight shift of the pump spot from the central axis of the stripe.

**Appendix 8: Polarization patterns without confinement**

Formation of the polarization patterns is determined to the greatest extent by the pseudospin evolution under the spin-orbit coupling. Neither a spatial confinement potential nor non-conservative processes lead to significant qualitative changes of the spatial distribution of the polarization. Let us consider the system described by the Hamiltonian $\hat{H}_\mathbf{k}$ coinciding with the Hamiltonian $\hat{T}$ in the main text:

$$\hat{H}_\mathbf{k} = \begin{bmatrix} \hbar^2 k^2/2m_{\text{eff}} & \Delta k^2 e^{-2i\theta} \\ \Delta k^2 e^{2i\theta} & \hbar^2 k^2/2m_{\text{eff}} \end{bmatrix}, \qquad (5)$$

where $q$ is the angle of injection characterizing the $\mathbf{k}$-state as $\mathbf{k} = (k\cos(q), k\sin(q))$. $\mathrm{D}$ is the splitting constant, $m_{\text{eff}}$ is the polariton effective mass. Here, we neglect both confinement and non-conservative processes. For simplicity, in this paragraph we omit the anisotropy splitting taking $d_{\text{an}} = 0$. The Hamiltonian (5) can be rewritten in terms of the effective magnetic field $\mathbf{\Omega}_\mathbf{k} = (W_{x,\mathbf{k}}, W_{y,\mathbf{k}}, 0)$ acting on the polariton pseudospin:

$$\hat{H}_\mathbf{k} = \frac{\hbar^2 k^2}{2m_{\text{eff}}}\hat{\sigma}_0 + \hbar\mathbf{\Omega}_\mathbf{k} \cdot \hat{\boldsymbol{\sigma}} = \frac{\hbar^2 k^2}{2m_{\text{eff}}}\hat{\sigma}_0 + \hbar(\Omega_{x,\mathbf{k}}\hat{\sigma}_x + \Omega_{y,\mathbf{k}}\hat{\sigma}_y), \qquad (6)$$

where the components of $\mathbf{\Omega}_\mathbf{k}$ are given as $W_{x,\mathbf{k}} = \mathrm{D}k^2\cos(2q)$, $W_{y,\mathbf{k}} = \mathrm{D}k^2\sin(2q)$. $\hat{\boldsymbol{\sigma}} = (\hat{\sigma}_x, \hat{\sigma}_y, \hat{\sigma}_z)$ is the vector of the Pauli matrices. The Hamiltonian $\hat{H}_\mathbf{k}$ yields to the following precession equation for the pseudospin vector $\mathbf{S}_\mathbf{k} = \langle\hat{\boldsymbol{\sigma}}\rangle = (S_{x,\mathbf{k}}, S_{y,\mathbf{k}}, S_{z,\mathbf{k}})$ around the effective magnetic field:

$$d_t S_k = \Omega_k \times S_k \qquad (7)$$

For the linear initial polarization, $S_{z0} \circ S_{z,k}(0) = 0$, the circular polarization evolution can be found from Eq. (7) analytically:

$$S_{z,k}(t) = \frac{1}{W}\left(W_{x,k} S_{y0} - W_{y,k} S_{x0}\right)\sin(Wt), \qquad (8)$$

where $W = |\Omega|$ is the Larmor frequency. From the general solution (8) of Eq. (7) we can find spatial distribution of the degree of the circular polarization (DOCP) along the sample, considering the polariton state with the wave vector $\mathbf{k}_1 = (k,0)$:

$$S_{z,\mathbf{k}_1}(t) = S_{y0}\sin(D k^2 t), \qquad (9)$$

which reduces to $S_{z,\mathbf{k}_1}(t) \approx S_{y0} D k^2 t$ close to the injection spot.

The solution (8) allows also to predict the polarization variation across the sample. We consider the case of a point-like source of polaritons, possessing the wave number $k$. Polaritons in the $\mathbf{k}$-state reach the crossection spaced by a distance $L_x$ from the injection spot in time of $T(q) = L_x/v_x \mu 1/k\cos(q)$. The resulting DOCP distribution in this crossection is found as

$$\begin{aligned}
S_{z,\mathbf{k}}(T(q)) &= \frac{1}{W}\left(W_{x,\mathbf{k}} S_{y0} - W_{y,\mathbf{k}} S_{x0}\right)\sin(WT(q)) \\
&\approx \left(W_{x,\mathbf{k}} S_{y0} - W_{y,\mathbf{k}} S_{x0}\right) T(q) \propto \frac{S_{y0}\cos(2q) - S_{x0}\sin(2q)}{\cos(q)}.
\end{aligned} \qquad (10)$$

Figure S8 allows to compare the numerical simulations and the analytical predictions of the spatial distribution of DOCP for different angles of the initial polarization plane. The panels show the simulated DOCP distributions (left). The right upper panels show the color bars illustrating the expected DOCP distribution across the sample found from Eq. (10). Approximate position of the crossection is labeled by the green dashed line in the left figure. The right lower panels show the diagram showing the dominating polarization in the plane of the sample calculated according to Eq. (9). The dash-dotted line in the left panel shows the examined direction. The dash-dotted line in the right lower figure indicates the circular polarization degree expected.

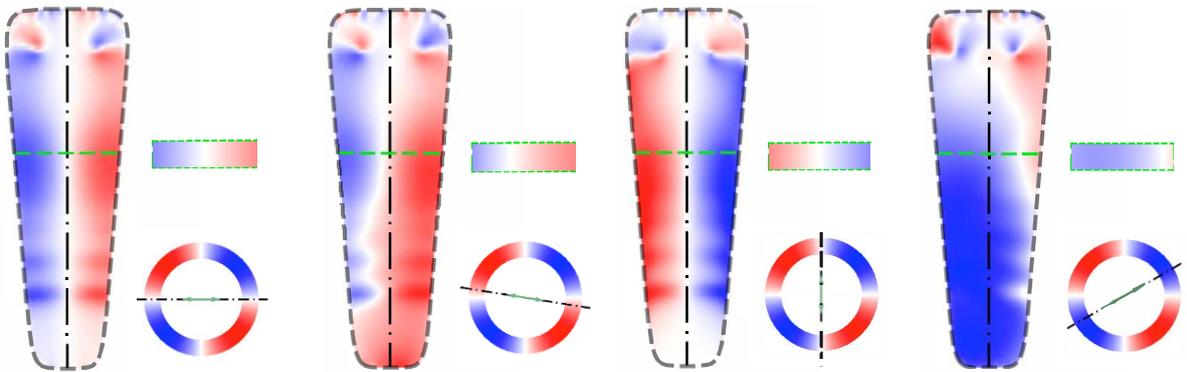

*Figure A8: Simulated and analytically calculated spatial distribution of the DOCP for various polarization orientations.*

**Appendix 9: Polariton pattern in a stipe**

We consider a stipe where polaritons can propagate along $y$ axis while their motion along $x$ axis is quantized. With account for the TE-TM splitting the effective Hamiltonian in the basis of the *linearly* polarized states reads

$$\mathcal{H} = \frac{\hbar^2 k^2}{2m}\hat{I} + \Delta\begin{pmatrix} k_x^2 - k_y^2 & 2k_x k_y \\ 2k_x k_y & k_y^2 - k_x^2 \end{pmatrix}. \tag{11}$$

Here $\hat{I}$ is the unit $2\times 2$ matrix, $m$ is the polariton effective mass, $\Delta$ is the TE-TM splitting parameter (which weakly depends on the polariton wavevector $k$). The Hamiltonian can be rewritten as

$$\mathcal{H} = \frac{\hbar^2}{2m}\begin{pmatrix} (1+\alpha)k_x^2 + (1-\alpha)k_y^2 & 2\alpha k_x k_y \\ 2k_x k_y & (1+\alpha)k_y^2 + (1-\alpha)k_x^2 \end{pmatrix}, \tag{12}$$

where $\alpha = 2m\Delta/\hbar^2$.

*Zero-order approximation.* Single subband model. We introduce the basic states

$$|x\rangle = \begin{pmatrix}1\\0\end{pmatrix}, \quad |y\rangle = \begin{pmatrix}0\\1\end{pmatrix},$$

and write the wavefunctions functions as (the normalization length is set to unity)

$$\Psi_x = e^{ik_y y}\cos\left(\frac{\pi x}{L}\right)|x\rangle, \quad \Psi_y = e^{ik_y y}\cos\left(\frac{\pi x}{L}\right)|y\rangle. \tag{13}$$

The dispersions of these states are given by

$$E_x(k_y) = \frac{\hbar^2 \pi^2}{2mL^2}(1+\alpha) + \frac{\hbar^2 k_y^2}{2m}(1-\alpha), \tag{14}$$

$$E_y(k_y) = \frac{\hbar^2 \pi^2}{2mL^2}(1-\alpha) + \frac{\hbar^2 k_y^2}{2m}(1+\alpha). \tag{15}$$

The TE-TM splitting of the states reads

$$\Delta_{TE-TM}(k) = \alpha\frac{\hbar^2}{m}\left(\frac{\pi^2}{L^2} - k_y^2\right), \tag{16}$$

and contains contributions due to the wavevector and due to the size-quantization.

The wavefunctions (13) correspond to definite linear polarization. In order to obtain the mixing of polarizations and circular polarization of polaritons we need to account for the off-diagonal terms in Eq. (12).

*First-order approximation.* Let us calculate the correction to the $\Psi_x$ due to the off-diagonal terms. The represent

$$\Psi_x = e^{ik_y y}\left[\cos\left(\frac{\pi x}{L}\right)|x\rangle + S(x)|y\rangle\right], \tag{17}$$

where $S(x)$ is the function which satisfies the equation

$$\left[\frac{\hbar^2}{2m}\frac{d^2}{dx^2} - E_x(k_y)\right]S(x) = -i\alpha\frac{\hbar^2 \pi k_y}{mL}\sin\left(\frac{\pi k_y}{L}\right). \tag{18}$$

The solution reads

$$S(x) = -\frac{i\alpha\hbar^2\pi k_y}{mL\left[\frac{\hbar^2\pi^2}{mL^2}-2E_x(k_y)\right]}\left[\sin\left(\frac{\pi x}{L}\right) - \frac{\sin\left(\sqrt{2mE_x(k_y)/\hbar}x\right)}{\sin\left(\sqrt{mE_x(k_y)/2\hbar}L\right)}\right]. \quad (19)$$

Neglecting $\alpha$ and $k_y$ dependence of $E_x(k_y)$ we have for the admixed function a much simpler expression

$$S(x) = -\alpha k_y x \cos\frac{\pi x}{L}. \quad (20)$$

The circular polarization degree in the state $\Psi_x$ is given, in first order in $\alpha$, by

$$P_c(x, k_y) = \frac{2S(x)\cos\frac{\pi x}{L}}{\cos^2\frac{\pi x}{L}} \approx -\alpha k_y x. \quad (21)$$

For the second state $\Psi_y$ the circular polarization degree $P_c(x, k_y) \approx \alpha k_y x$.